\documentstyle[twoside,fleqn,espcrc2]{article}

\newcommand{\AmS}{{\protect\the\textfont2
  A\kern-.1667em\lower.5ex\hbox{M}\kern-.125emS}}

\title{Deconfinement from action restriction}

\author{M. Grady\address{Department of Physics, SUNY Fredonia, 
Fredonia NY 14063 USA}}%
       
\begin{document}

\begin{abstract}
The effect of restricting the plaquette to be greater than a certain cutoff
value is studied. The action considered is the standard Wilson action with
the addition of a plaquette restriction, 
which should not affect the continuum limit
of the theory. In this investigation, the strong coupling limit
is also taken. It is found
that a deconfining phase transition occurs as the cutoff is increased, on
all lattices studied (up to $20^4$). The critical cutoff on the infinite
lattice appears to be around 0.55. For cutoffs above this, a fixed point behavior
is observed in the normalized fourth cumulant of the Polyakov loop, suggesting
the existence of a line of critical points corresponding to a massless gluon
phase, not unlike the situation in compact U(1). The Polyakov loop susceptibility also
appears to be diverging with lattice size at these cutoffs. A
strong finite volume behavior is observed in the 
pseudo-specific heat. It is discussed
whether these results could still be consistent with the standard crossover
picture which precludes the existence of a deconfining phase transition on 
an infinite symmetric lattice.

\end{abstract}

\maketitle

\section{INTRODUCTION}

In SU(2) lattice gauge theory the plaquette, 
$P \equiv \frac{1}{2} \mbox{tr} U_{\Box}$ takes values from $-1$ to $1$.
The action studied is  $S_{\Box} = S_W = (1-P)$ if $P \ge c$ 
and $\infty$ if $P < c$, where
$c =$ cutoff parameter, $-1 \le c \le 1$.  For instance, the 
positive plaquette action\cite{mack} has $c=0$.  
Since the continuum limit is determined only by the behavior of the 
action in an infinitesimal region around its minimum which occurs around $P=1$,
this action should have the same continuum limit as the Wilson action
for all values of $c$ except $c=1$.
This study was primarily designed to determine whether a sufficiently
strong cutoff would cause the system to deconfine. To give
the lattices the maximum chance to confine, the strong coupling limit,
$\beta \rightarrow 0^+$, was taken. This removes the Wilson
part of the action completely, leaving only the plaquette restriction.
The study was carried out on symmetric lattices from $6^4$ to $20^4$ in an
attempt to study the zero-temperature system. However, as is well known, 
symmetric lattices are not really at zero temperature until one takes the
limit of infinite lattice size. Finite size lattices are 
at a finite temperature corresponding to their inverse time extent. In this
sense, finite symmetric lattices can be thought of as small spatial-volume
finite temperature systems. The big question is whether the transition observed
here as a function of the cutoff is a bulk transition which will remain
on the infinite lattice, or the remnant of the usual finite temperature 
transition which will disappear by moving to $c=1$ on the infinite lattice.

\section{MOTIVATION}
The purpose of an action restriction is to eliminate lattice 
artifacts which occur at strong coupling, and which may give misleading
results that have nothing to do with the continuum limit.  For instance,
the original motivation of the positive plaquette action was to eliminate
single-plaquette
$Z_2$ monopoles and strings, which were possibly thought to be responsible 
for confinement \cite{mack,grady1}. However, 
for strong enough couplings the positive-plaquette
action was shown to still confine \cite{heller}. Eliminating these artifacts, does,
however,
appear to remove the ``dip'' from the beta-function in the crossover region,
possibly improving the scaling properties of the theory, and suggesting that 
this dip is indeed an artifact of non-universal 
strong coupling aspects of the Wilson action.

For the U(1) theory, all monopoles are eliminated for $c \ge 0.5$, since
in this case six plaquettes cannot carry a visible flux totaling $\pm 2 \pi$,
($\cos (\pi /3) = 0.5$). Therefore a Dirac string cannot end and there are no
monopoles. The theory will be deconfined for all $\beta$ for $c\ge 0.5$, 
since confinement is due to
single-plaquette monopoles in this theory.
SU(2) confinement may be related to U(1) confinement through abelian
projection. It is an interesting question whether a similar 
action restriction which limits the non-abelian flux carried by an SU(2) 
plaquette will eliminate the associated abelian monopoles causing
deconfinement, or whether
surviving large monopoles can keep
SU(2) confining	for all couplings. A related question is how can 
the SU(2) average plaquette go to unity in
the weak coupling limit, but the corresponding effective U(1) 
average plaquette stay in the 
strong coupling confining region ($\leq .6$)? Chernodub et.\ al.\ have found
$<\!\! P_{\rm{U(1)}}\!\! > \geq 0.82<\!\! P_{\rm{SU(2)}}\!\! >$ 
over a wide range of couplings \cite{ch}.
\begin{figure}[htb]
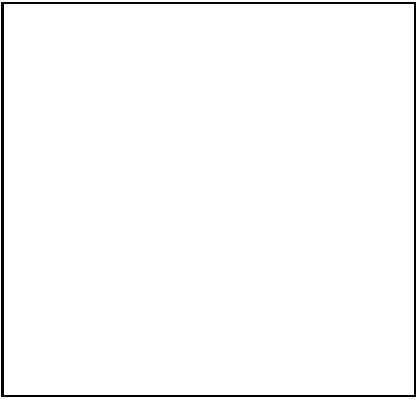

\vspace{9pt}
\framebox[55mm]{\rule[-21mm]{0mm}{50mm}}
\caption{Polyakov Loop Modulus. The lower curves are reduced by the 
value that a Gaussian of the same width would be expected to have.}
\label{fig:2}
\end{figure}
\begin{figure}[htb]
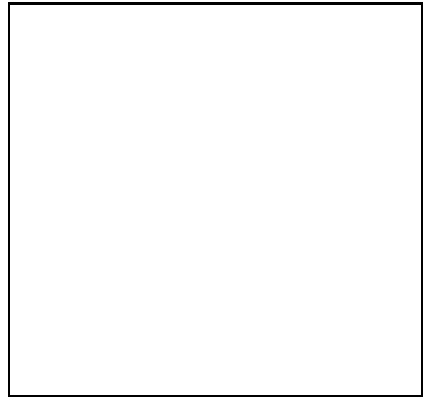

\vspace{9pt}
\framebox[55mm]{\rule[-21mm]{0mm}{50mm}}
\caption{Polyakov loop histograms from $8^4$ lattice showing typical symmetry breaking behavior.}
\label{fig:3}
\end{figure}
\section{Results}
A deconfining phase transition is observed on all lattices studied 
($6^4$ to $20^4$). The behavior
of the Polyakov loop, $L$, shows the normal symmetry breaking behavior and
is smooth, suggesting a second or higher order transition (Figs 1,2).
The transition point has a clear lattice size (N) dependence (Figs 1,3).
The normalized fourth cumulant, $g_4 \equiv 3 - \!\! <\!\! L^4 \!\! >\!\! /\!\! 
< \!\! L^2 \!\! >^2  $,
shows fixed point behavior 
(no discernible lattice size dependence) for $c\geq 0.6$ 
at a non-trivial value around $g_4 = 1.6$ (Fig 3). 
\begin{figure}[htb]
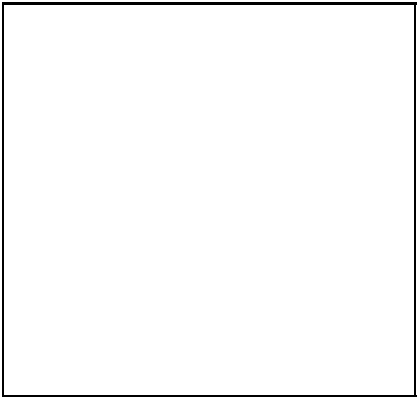

\vspace{9pt}
\framebox[55mm]{\rule[-21mm]{0mm}{50mm}}
\caption{Normalized fourth cumulant of the Polyakov Loop.}
\label{fig:4}
\end{figure}
This suggests either
a line of critical points for $c \ge 0.6$, e.g. from a massless gluon phase,
or that correlation lengths are so 
large that finite lattice size dependence is hidden. For the standard 
interpretation to
hold the 
normalized fourth cumulant should go to zero as lattice size approaches 
infinity
for all values of $c$.
The Polyakov loop susceptibility shows signs of diverging with lattice size ($N$) for 
$c \geq 0.6$ and converging to a finite value for $c \leq 0.5$ (Fig. 4). This supports
the idea of a different behavior in these regions in the infinite volume
limit.
\begin{figure}[htb]
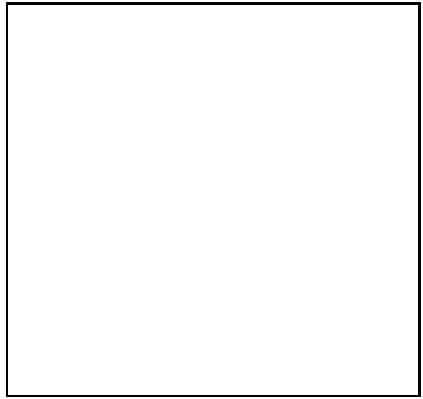

\vspace{9pt}
\framebox[55mm]{\rule[-21mm]{0mm}{50mm}}
\caption{Lattice size dependence of the plaquette susceptibility. Constant
large-lattice behavior for lower cutoffs indicates that dividing by $<\! L\! >$
correctly normalizes the changing order parameter (the definition changes
with lattice size).}
\label{fig:6}
\end{figure}
Extrapolations of finite lattice ``critical points'' to infinite lattice
size are consistent with infinite lattice critical points ranging from
around $c^{\infty}=0.55$ to $c^{\infty}=1.0$ depending on scaling function assumed,
so are basically inconclusive (Fig 5). Although a straightforward extrapolation
gives infinite lattice critical cutoff around $c^{\infty}=0.55$, a logarithmic scaling
function can be made consistent with $c^{\infty} = 1.0$ in which case the transition
would no longer exist on the infinite lattice.
Finally, the pseudo-specific heat from plaquette fluctuations shows a 
definite signal
of a broad peak or shoulder beginning to form for the 
largest lattice sizes (Fig 6), 
near
the location of deconfinement. Interestingly, no noticeable effect 
was seen here on the smaller lattices, but the $16^4$ and $20^4$ results are
definitely shifted up. This indicates a rather large
scaling exponent. Preliminary analysis shows an exponent near $4$, i.e.
a peak height diverging according to the space-time volume (if indeed
it is a peak). This would indicate a possible weak first-order transition.
Since this is a bulk (local in space-time) quantity, a divergence here would
be a surprise for a finite-temperature transition.
Clearly larger lattices and better statistics are needed
to get the full picture behind this rich and interesting system.
\begin{figure}[htb]
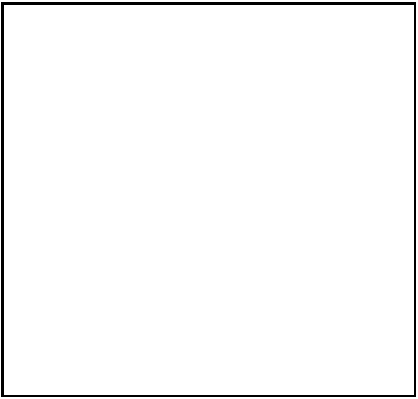

\vspace{9pt}
\framebox[55mm]{\rule[-21mm]{0mm}{50mm}}
\caption{Extrapolation of critical point to infinite volume. The finite 
lattice critical point is arbitrarily defined 
by $g_4 (c^*)=0.25.$}
\label{fig:7}
\end{figure}
\begin{figure}[htb]
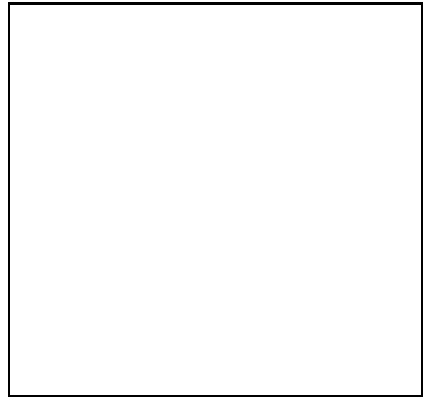

\vspace{9pt}
\framebox[55mm]{\rule[-21mm]{0mm}{50mm}}
\caption{Pseudo-specific heat from plaquette fluctuations.}
\label{fig:8}
\end{figure}
\section{CONCLUSIONS}
The imposition of an action restriction clearly induces 
a deconfining ``phase transition'' on a finite symmetric 
lattice,
even in the strong coupling limit (of course there really is no phase transition on 
a finite lattice).
The extrapolation to infinite lattice size is, as always, difficult. There are two
possibilities.
Evidence is consistent with deconfinement on the infinite lattice
for all $c \! > \! 0.55$, representing possibly a massless gluon phase in the 
continuum limit, similar to the U(1) continuum limit. It should be 
remembered that the compact U(1) theory itself undergoes a similar deconfining
transition at $c=0.5$. The mechanism in both cases may be the same,
namely that limiting the flux that can be carried by an individual plaquette
prevents a Dirac string from ending, and spilling its flux out in all
directions in the form of  a point monopole. Although it has proven
hard to identify monopoles in the SU(2) theory, the identification of them
with abelian monopoles appearing in the Maximal Abelian Gauge suggests
a possibly similar behavior in SU(2) and U(1).

It is also possible, however, that all signs of critical behavior will
disappear on large enough lattices, nothing will diverge, and the theory
will confine for all values of c, (i.e. $c^{\infty} = 1.0$), the apparent
critical behavior being due to the finite temperature associated with
finite $N$. However, as a small 3-volume 
usually {\em masks} critical behavior by smoothing would-be criticalities
it is difficult to see how this could explain the large divergent-looking 
finite size effects 
being seen here. It is especially difficult to explain the large-lattice
behavior of the pseudo-specific heat.
Due to the small
spatial volume and the fact that one is very far from the scaling region,
however, it is rather difficult to make definite predictions
from this scenario.

One also must remember that our parameter is the cutoff, $c$, and 
not the coupling
constant. Although $c$ does seem to play the role of an effective coupling,
(with $c=-1$ corresponding to $\beta=0$ in the Wilson theory, and $c=1$ corresponding
to $\beta=\infty$), it certainly differs in some respects. Incidentally,
the average plaquette at $c=0.5$ is 0.7449, corresponding to a Wilson
$\beta$ of around 3.2.  This relatively weak coupling means that rather
large lattices will be needed to disentangle the finite temperature effects,
from any possible underlying bulk transition.

It is likely that further light could be shed on this subject by 
gauge transforming configurations from restricted action
simulations to the maximum abelian gauge and extracting abelian 
monopole loops.  
This would answer the question of what effect the SU(2) action restriction
has on the corresponding abelian monopoles.
By studying the strength of the monopole suppression it may be possible 
to tell whether a sufficient density of monopoles will survive the
continuum limit to produce a confining theory. 
Such a study is underway.

It is clear that gauge configurations from the restricted action will be 
relatively smoother on the smallest scale. It should be pointed out, however, that
anything can (and does) still happen at larger scales.  For instance with $c=0.5$,
even the $2\times 2$ Wilson loop can take on any value. The maximum flux
carried by a plaquette is only limited to roughly $1/3$ of its unrestricted value,
i.e. around the value expected on a lattice twice as fine as 
that in a typical unrestricted simulation. From this point of view, one 
should only
need lattices twice as large in linear extent to see the same physics in
a $c=.5$ restricted simulation as in an unrestricted simulation.
The smoother configurations produced by restrictions of this strength
may also eliminate or at least suppress dislocations, possibly 
enabling one to study
instantons without the need for cooling.

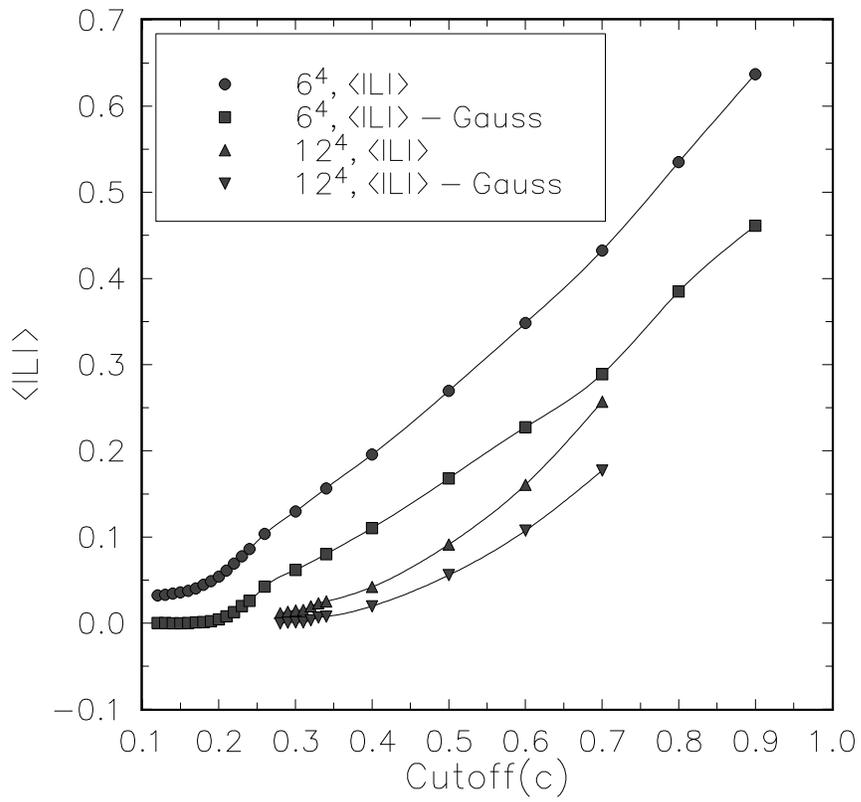
\end{document}